\documentclass[preprint2,times,tighten]{aastex6}

\usepackage{color}
\usepackage{enumitem}
\usepackage{hyperref}
\usepackage{verbatim}
\usepackage{amsmath,amstext,mathrsfs}
\usepackage[all]{hypcap} 
\usepackage{afterpage}

\begin{document}
\title{Tracing interstellar magnetic field using velocity gradient technique:\\ Application to Atomic Hydrogen data}

\author{Ka Ho Yuen\altaffilmark{1,2}, A. Lazarian\altaffilmark{1}}
\email{kyuen2@wisc.edu, lazarian@astro.wisc.edu}
\altaffiltext{1}{Department of Astronomy, University of Wisconsin-Madison}
\altaffiltext{2}{Department of Physics, The Chinese University of Hong Kong}
\begin{abstract}
The advancement of our understanding of MHD turbulence opens ways to develop new techniques to probe magnetic fields. In MHD turbulence, the velocity gradients are expected to be perpendicular to magnetic fields and this fact was used by \cite{2016arXiv160806867G} to introduce a new technique to trace magnetic fields using velocity centroid gradients. The latter can be obtained from spectroscopic observations. We apply the technique to GALFA HI survey data and compare the directions of magnetic fields obtained with our technique with the direction of magnetic fields obtained using PLANCK polarization. We find excellent correspondence between the two ways of magnetic field tracing, which is obvious via visual comparison and through measuring of the statistics of magnetic field fluctuations obtained with the polarization data and our technique. This suggests that the velocity centroid gradients has a potential for measuring of the foreground magnetic field fluctuations and thus provide a new way of separating foreground and CMB polarization signals. 
\end{abstract}

\keywords{ISM: general --- ISM: structure --- magnetohydrodynamics (MHD) --- radio lines: ISM --- turbulence}
\section{Introduction}

Turbulence is ubiquitous in astrophysics. The Big Power Law in the Sky \citep{Armstrong1995ElectronMedium,Chepurnov2010ExtendingData} shows clear evidence that interstellar turbulence extends over 10 orders of magnitude of scales in the interstellar media (ISM). The ISM is magnetized and therefore the turbulence is magnetohydrodynamic (MHD) in nature, e.g. see \citep{Li2014a,Zhang2014,Pillai2015}.

The modern theory of turbulence has been developed on the basis of the prophetic work by \citeauthor{GoldreichP.Sridhar1995GS95IITurbulence}, (\citeyear{GoldreichP.Sridhar1995GS95IITurbulence}, henceforth GS95).  The original ideas were modified and augmented in subsequent theoretical and numerical studies (\citealt{Lazarian1999ReconnectionField, Cho2000TheTurbulence,Maron2000SimulationsTurbulence,Lithwick2001CompressiblePlasmas,Cho2001SimulationsMedium,Cho2002CompressiblePlasmasb,Cho2003CompressibleImplicationsb,Kowal2010VelocityScalingsb}, see \citealt{Brandenburg2013AstrophysicalTurbulence} for a a review).\footnote{We do not consider the modifications of the GS95 model that were intended to explain the spectrum $k^{-3/2}$ that was reported in some numerical studies (e.g. \cite{Boldyrev2006SpectrumTurbulence}). We believe that the reason for the deviations from the GS95 predictions is the numerical bottleneck effect, which is more extended in the MHD compared to hydro turbulence \citep{Beresnyak2010ScalingTurbulence}. This explanation is supported by high resolution numerical simulations that correspond to GS95 predictions (see \cite{Beresnyak2014SpectraSimulations}). The simulations also strongly support the anisotropy predicted in GS95 and rule out the anisotropy prediction in the aforementioned alternative model.} The Alfvenic incompressible motions dominate the cascade. This cascade can be visualized as a cascade of elongated eddies rotating perpendicular to the {\it local} direction of the field.\footnote{The notion of the local direction was not a part of the original GS95 model. It was introduced and justified in more recent publications (see \citealt{Lazarian1999ReconnectionField, Cho2000TheTurbulence,Maron2000SimulationsTurbulence}).} Naturally, this induces the strongest gradients of velocity perpendicular to the magnetic field. Thus one can expect that measuring the gradient in turbulent media can reveal the local direction of magnetic field. This property of velocity gradients was employed in \citeauthor{2016arXiv160806867G} (\citeyear{2016arXiv160806867G}, hereafter GL16) to introduce a radically new way of tracing magnetic fields using spectroscopic data. Instead of using aligned grains or synchrotron polarization (see \citealt{Draine2011PhysicsMedium}), GL16 applied velocity centroid gradients (henceforth VCGs) to synthetic maps obtained via MHD simulations and obtained a good agreement between the projected magnetic fields and the directions traced by the VCGs. As the velocity centroids can be readily available from spectroscopic observations (see Esquivel \& Lazarian 2005), this provided a way not only for observational tracing of magnetic fields but also for finding its strength using the GL16 technique that is similar to the well-known Chardrasechar-Fermi method.   

Motivated by the GL16 study, in this paper we calculate the VCGs using HI data from the GALFA survey\citep{Peek2011The1} and compare the directions of the magnetic fields that we trace using the gradients with the directions of magnetic fields that are available from the PLANCK polarization survey \citep{Adam2016iPlanck/iResults}.\footnote{Based on observations obtained with Planck (http://www.esa.int/Planck), an ESA science mission with instruments and contributions directly funded by ESA Member States, NASA, and Canada.} To do this, we first significantly improve the procedure of calculating of the VCGs and test it with numerical data. Our recipe for calculating the VCGs is presented in \S \ref{sec:recipe}, while in \S \ref{sec:Observations}, we apply the technique to trace magnetic fields.  We discuss our results in \S \ref{sec:discussions}, and our conclusions are presented in \S \ref{sec:conclusions}. 

\section{Improved procedure for calculating velocity gradients}
\label{sec:recipe}

GL16 established that the VCGs can trace magnetic field in MHD turbulence. However, this exploratory study lacks a criterion on judging on how well gradients can trace magnetic fields. Therefore it is difficult to judge what is the resolution requirement to trace magnetic field vectors and and what are the uncertainties. Therefore our first goal is to introduce a more robust procedure of the VCGs calculation which is to return the tracing that is independent on the resolution of the simulations and only depends
on the parameters of MHD turbulence. 

We used a single fluid, operator-split, staggered grid MHD Eulerian code ZEUS-MP/HK,\footnote{Maintained by Otto \& Yuen, (https://bitbucket.org/cuhksfg/zeusmp-hk/)} a variant of the well-tested code ZEUS-MP \citep{Norman2000,Hayes2006}, to set up a three-dimensional, uniform, isothermal, supersonic, sub-Alfvenic turbulent medium. We adopted periodic boundary conditions. The initial cube was set with a uniform density, and an initial uniform field. Turbulence was injected solenoidally continuously, e.g. see \citep{Ostriker2000}, see also Appendix of \cite{Otto17}. Our simulations had the resolution of $792^3$. We selected two cubes with sonic Mach number $M_s=5$ and Alfvenic Mach number $M_A=0.6$ but different initial magnetic field orientation (one was parallel to the $z$-axis, another is at the angle $\pi/7$ to the $z$-axis). Compared to the GL16, we used higher resolution simulations and studied the effect of varying magnetic-field direction relative to the line of sight. 

To trace magnetic field we generated polarization maps by projecting our data cubes along the $x$-axis and assuming that the dust producing the polarization followed the gas and was perfectly aligned by the magnetic field. Let $\phi=\tan^{-1}(B_y/B_z)$, where $B_{y,z}$ are the y and z direction of magnetic field. The intensity $I$, velocity centroid $C$ and stokes parameters $Q$, $U$ were computed by :
\begin{equation}
\begin{aligned}
I({\bf r}) &= \int \rho({\bf r},x)  dx\\
C({\bf r}) &=I^{-1} \int \rho({\bf r},x) v_x({\bf r},x)  dx\\
Q({\bf r}) &\propto \int \rho({\bf r},x) \cos{2\phi}  dx\\
U({\bf r}) &\propto \int \rho({\bf r},x) \sin{2\phi}  dx\\\\
\end{aligned}
\end{equation}
where ${\bf r}$ is the vector on the $y-z$ plane. The polarization angle is given by $\phi_{2d} = 0.5\tan^{-1}(U/Q)$. Polarization  traces the magnetic field projected along the line of sight.

We calculated velocity centroids following GL16 but modified the VCGs calculations to increase the accuracy of the procedure. In particular, we performed cubic spline interpolation, which uses a three-point estimate to provide the maps for gradient study. The resulting map is 10 times larger than the original one. To search for maximum gradient direction in each data point, we selected a  neighborhood of the radius vector $r\in (0.9,1.1)$ pixels in the interpolated map. The interpolation process is accurate with a $3^{o}$ error, and is comparable to the Sober operator used in  \cite{Soler2013}. We smoothed our data with a $\sigma=2$ pixels Gaussian kernel. 
\begin{figure}[t]
\centering
\includegraphics[width=0.44\textwidth]{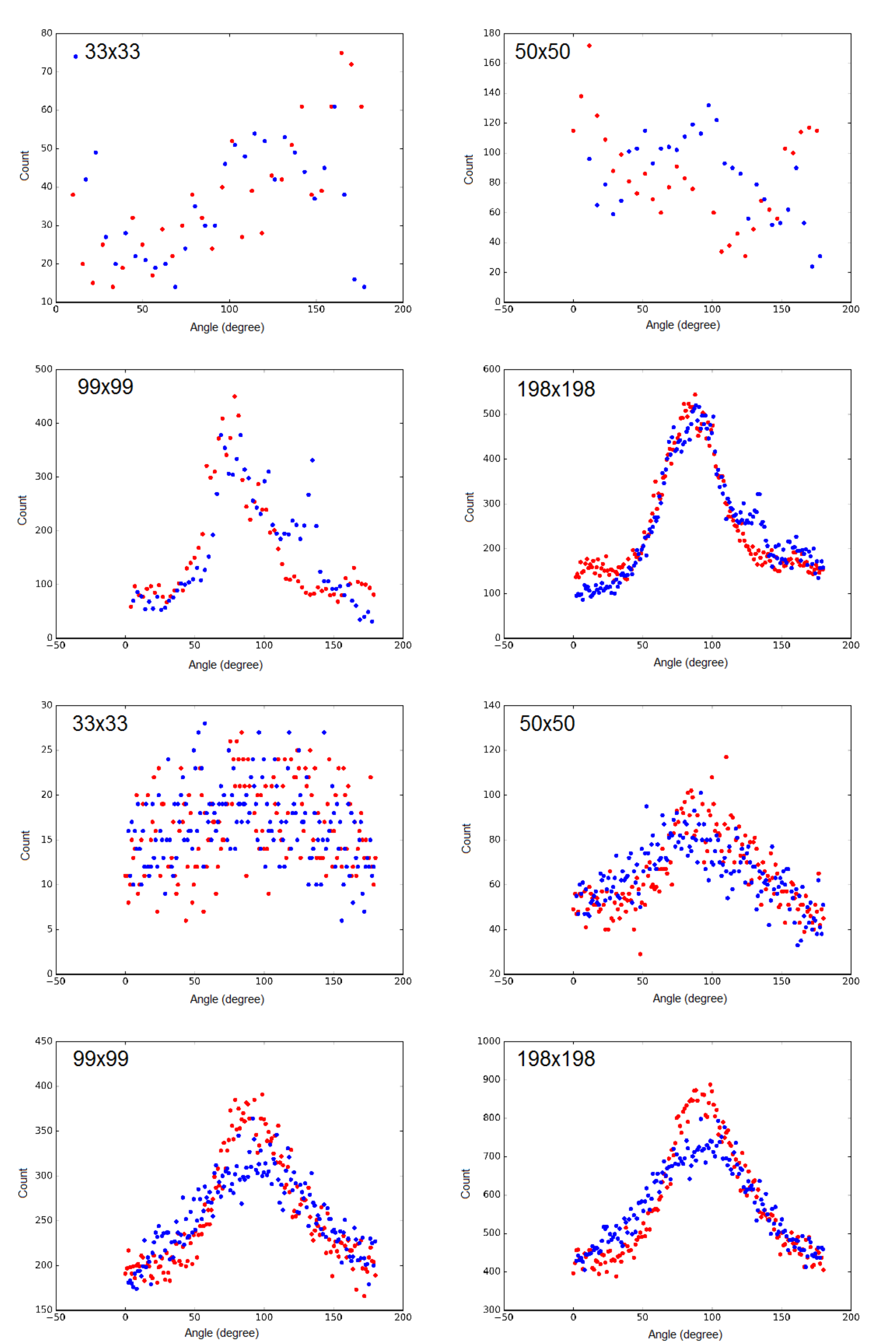}
\caption{\label{fig:AARA} (Upper four) The distribution of absolute angle (red) and relative angle (blue) in a synthetic map of size 792x792 for sub-regions of size 33x33, 50x50, 99x99, 198x198, respectively. The Gaussian profile emerges when the patch is 1/8 of the total length of the map. The profile is well-defined when it is 1/4 of the map. (Lower four) The distributions of absolute angle (red) and relative angle (blue) from observation data for sub-region of size 50x50, 100x100, 200x200, 300x300 (relative to GALFA-HI data resolution) respectively. }
\end{figure}

\afterpage{\pagebreak}
\begin{figure*}[pt!]
\centering
\includegraphics[width=0.88\textwidth]{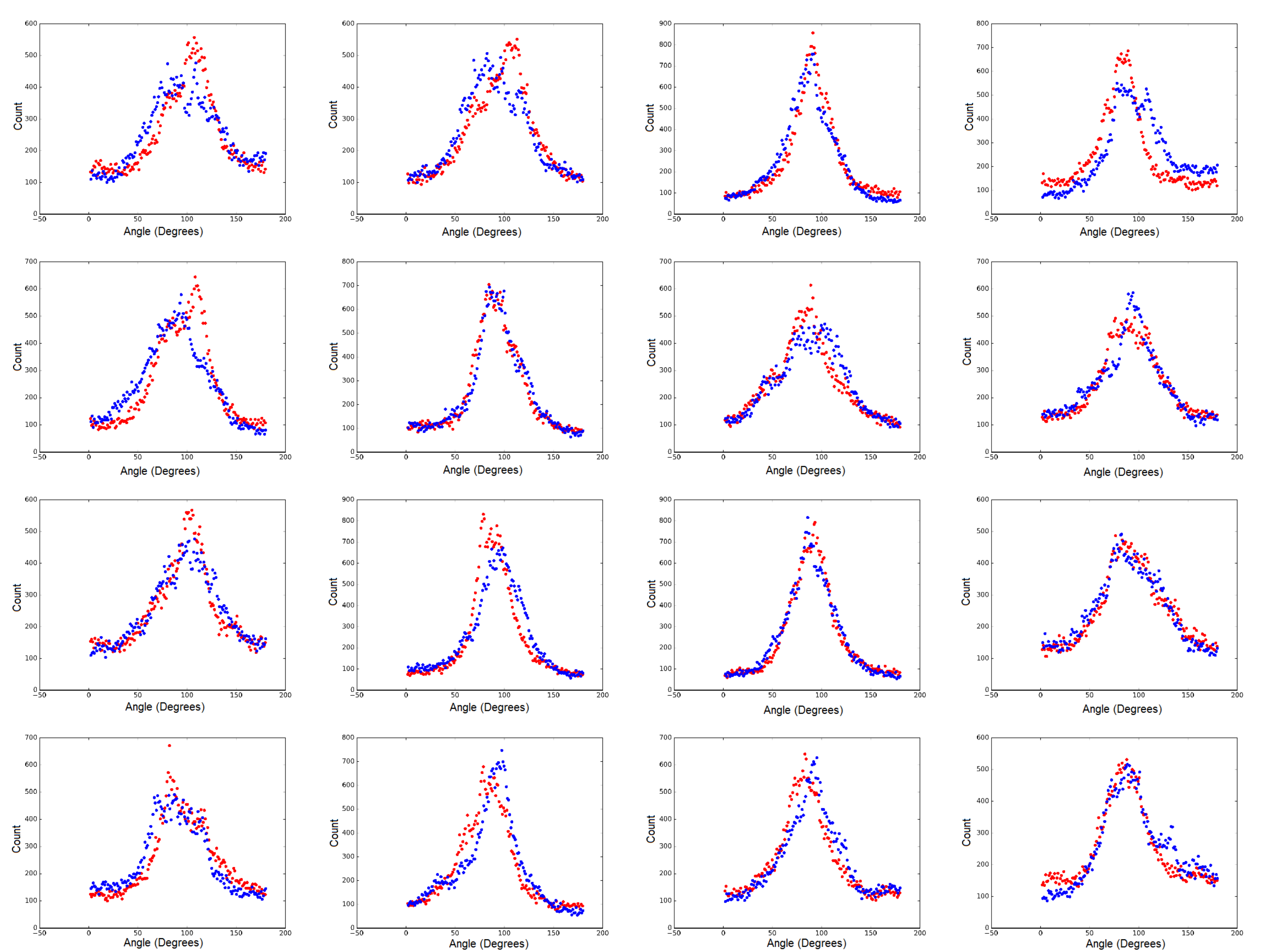}
\includegraphics[width=0.60\textwidth]{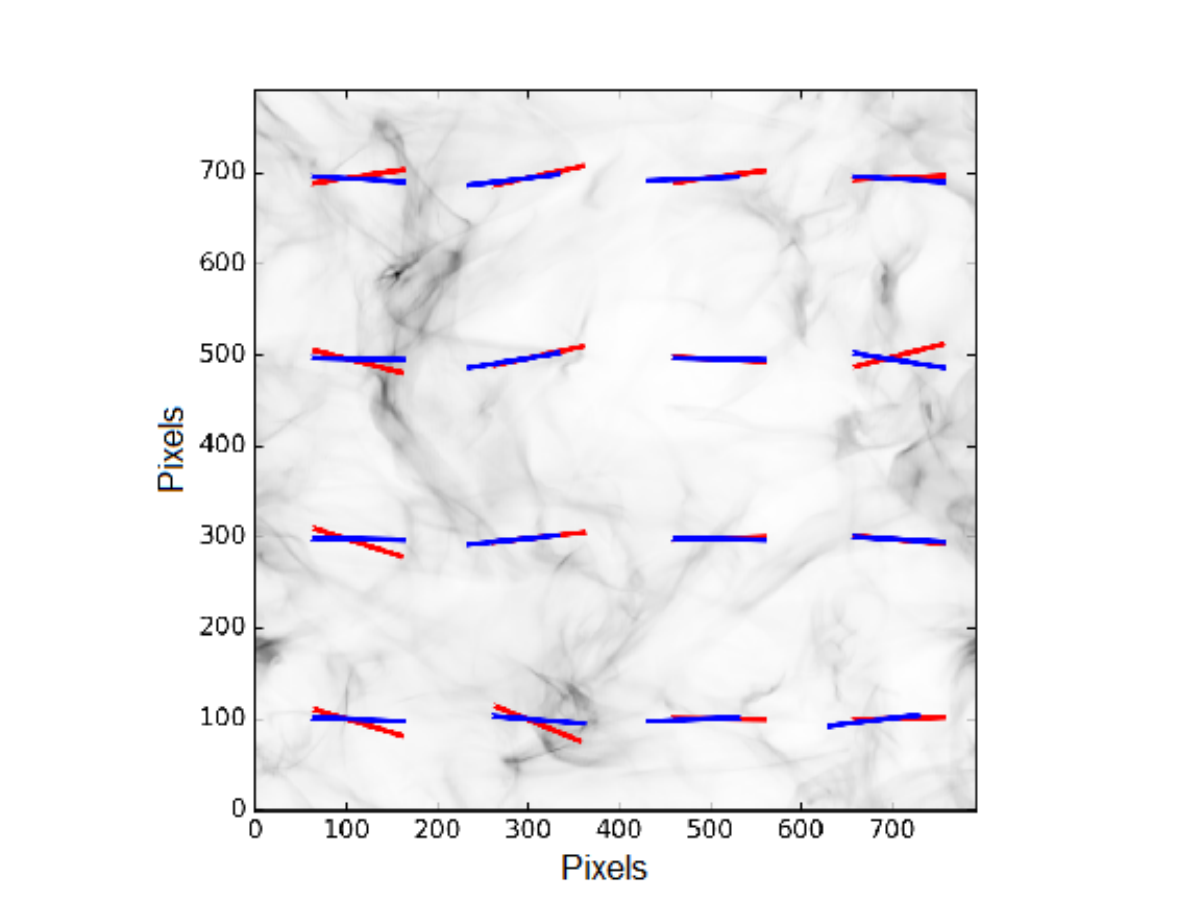}
\caption{\label{fig:gaussianvectors} (Upper 16 panels) The distribution of AA (red) and RA (blue) in a synthetic map from {\it run-2} with block size $198\times198$.  By detecting the peak of the AA distribution, we determined the mean magnetic field direction within the block. 
(Lower) The predicted mean magnetic field vector (red) compared with the real magnetic field vector(blue). The background is the intensity of the synthetic map.}

\end{figure*}

The statistical properties of gradient fields can determine the mean direction of magnetic fields in a sub-region of interest. We divided our synthetic maps into sub-regions and examined the statistical behavior of gradient vector orientation (hereafter absolute angle (AA)) and relative angle $\phi$ between gradients and fields (hereafter relative angle (RA)) within the region.  The upper four panels of \ref{fig:AARA} shows what distributions of the AA and RA look like when size of the block decreases. As the block size increases, the mean gradient direction becomes more well-defined. The alignment between the gradient and magnetic field also becomes more clear as block size increases. We find that as the block size arrives at $~100\times100$, a sharp distribution emerges with well-defined mean and dispersion. By measuring the mean of the AA distributions, we determine the mean magnetic field direction within the respective block. The RA distributions tells us how accurate this prediction of magnetic fields is. We shall call this treatment {\it sub-block averaging} in the following sections. Notice that, sub-block averaging is not a smoothing method. It is used to increase the emphasis of important statistics  and suppress noise in a region, and provide an estimate on how accurate this averaging is by the AA-RA diagram. On the other hand, smoothing does not provide such an estimate. A detailed discussion of how white noise affects the sub-block averaging and smoothing is provided in an extended paper by Lazarian et al. (2017), where the a companion new measure, namely, synchrotron intensity gradients are studied. 

The benefits of our approach can be seen in Figure \ref{fig:gaussianvectors}. We divided the whole simulation domain into 16 blocks with equal size, and predicted the magnetic field direction in each block.  As one can see from these figures, the VCGs trace well magnetic fields. We also confirmed this for synthetic observations when the line of sight was at different angles to the mean direction of magnetic field.  

 \citeauthor{Chandrasekhar1953ProblemsField.} (\citeyear{Chandrasekhar1953ProblemsField.}, C-F) provides an expression relating the strength of of plane-of-sky magnetic field by dispersion of turbulent velocities $\delta v$ and polarization vectors $\delta \theta$ in magnetized turbulence (For an improved C-F method, see \citealt{Falceta-Goncalves2008StudiesTechnique}):
\begin{equation}
\begin{aligned}
\delta B \sim \sqrt{4\pi\rho} \frac{\delta v}{\delta \phi}
\end{aligned}
\end{equation}

The mean magnetic field strength can also be calculated using the same concept in sub-block averaging.The dispersion of VCGs and that of magnetic-field directions are not exactly the same, but the difference is small. GL16 introduced a factor $\gamma$ of $\sim 1.29$ to account for this difference. In our case, using our improved procedure of gradient calculation we get the dispersion of the VCGs in blocks that is just 1.07-times that of polarization. The standard deviation of the ratio of the dispersions is 0.05.  As illustrated in GL16, the factor $\gamma$ varies with parameters of MHD turbulence. Elsewhere we shall provide a fitting expression for $\gamma$ as the function of $M_s$ and $M_A$. This should further increase the accuracy of obtaining the value of magnetic field strength. More details on the technique of obtaining magnetic field intensity using only spectroscopic information and no polarimetry will be provided in our forthcoming paper (Yuen \& Lazarian, in preparation).

\section{Application to observation data}
\label{sec:Observations}
\begin{figure*}[t]
\centering
\includegraphics[width=0.98\textwidth]{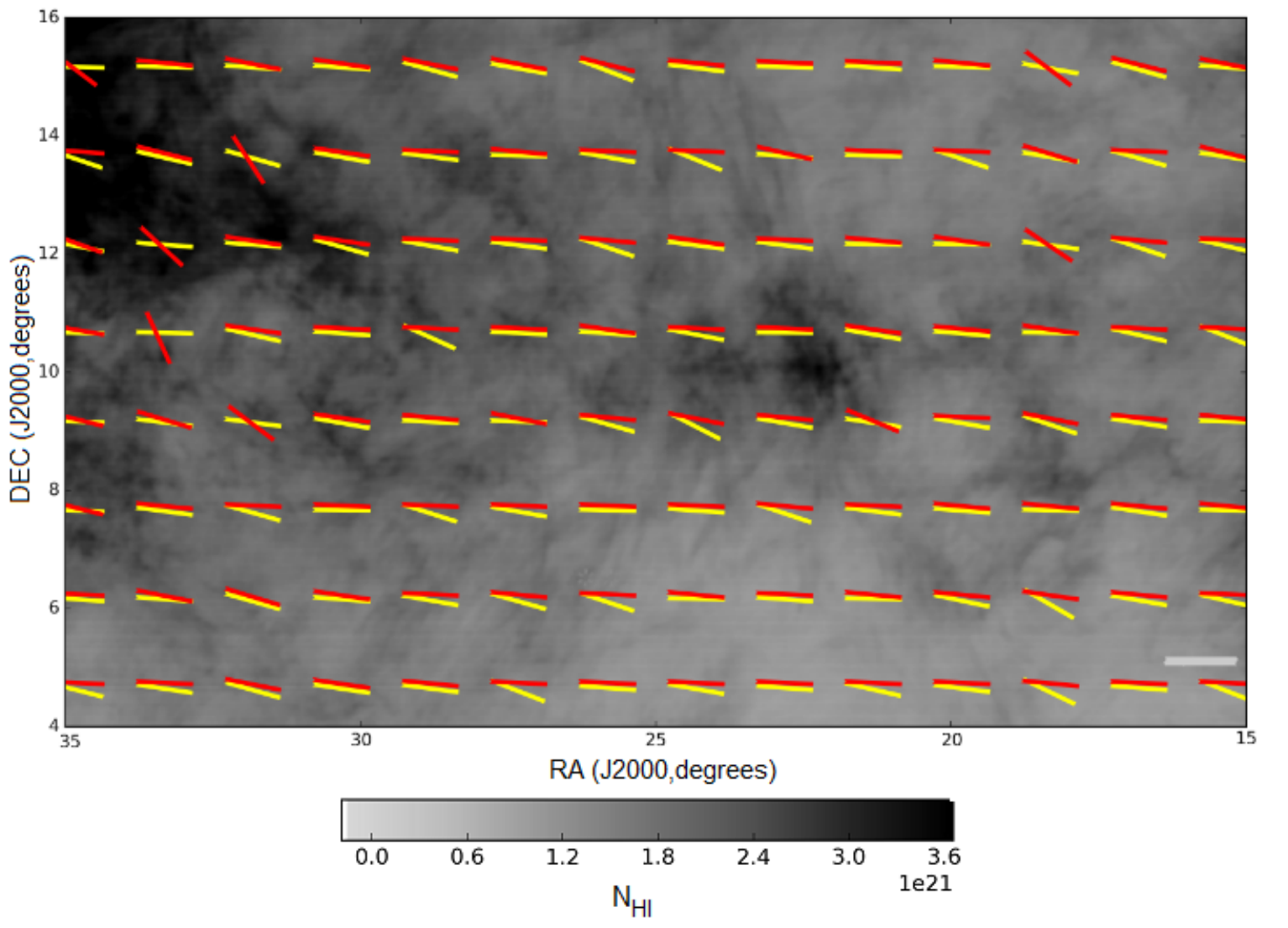}
\caption{\label{fig:gadfi-hi-vgbb}{\it Rotated} VCGs (Yellow) map obtained using GALFA-HI data. Red vectors are polarization directions obtained from the {\it PLANCK} data. The directions presented in this figure show the direction but not the magnitude. The background shows the column density of atomic hydrogen.}
\end{figure*}

With the tested procedure in hand,  we selected diffuse regions from observation surveys. We acquired data from the Galactic Arecibo L-Band Feed Array HI Survey (GALFA-HI). We compare the VCGs directions to the {\it PLANCK} polarization data. In diffuse media, polarization of emitted radiation is perpendicular to local magnetic field direction \citep{2007JQSRT.106..225L,Andersson2015InterstellarAlignment}, i.e. the same way as the VCGs. To adapt the difference of resolutions, we adjust the block size used in Planck to reflect the same physical block GALFA is referring to.

The region we selected from GALFA-HI survey data spans right ascension $15^o$ to $35^o$ and declination $4^o$ to $16^o$. The bin size along the velocity axis is 0.18 km/s.  We analyzed 353GHz polarization data obtained by the {\it Planck} satellite's High Frequency Instrument (HFI).\footnote{We use the {\it planckpy} module to extract polarization data in a particular region with J2000 equatorial coordinate: (https://bitbucket.org/ezbc/planckpy/src)} We performed the same procedure as indicated in Section \ref{sec:recipe}. We checked the AA and RA, as shown in the lower 4 panels of \ref{fig:AARA}, to pick an appropriate block size for a gradient vector. For the given case, a $100\times 100$ block satisfies the requirement in the recipe. The velocity gradient vectors are plotted with polarization vectors in Figure \ref{fig:gadfi-hi-vgbb}. In this region, most of the gradient vectors align very well with polarization vectors. The detailed study of the observed deviations from the perfect alignment will be provided in our subsequent publication.

\begin{figure}[t]
\centering
\includegraphics[width=0.48\textwidth]{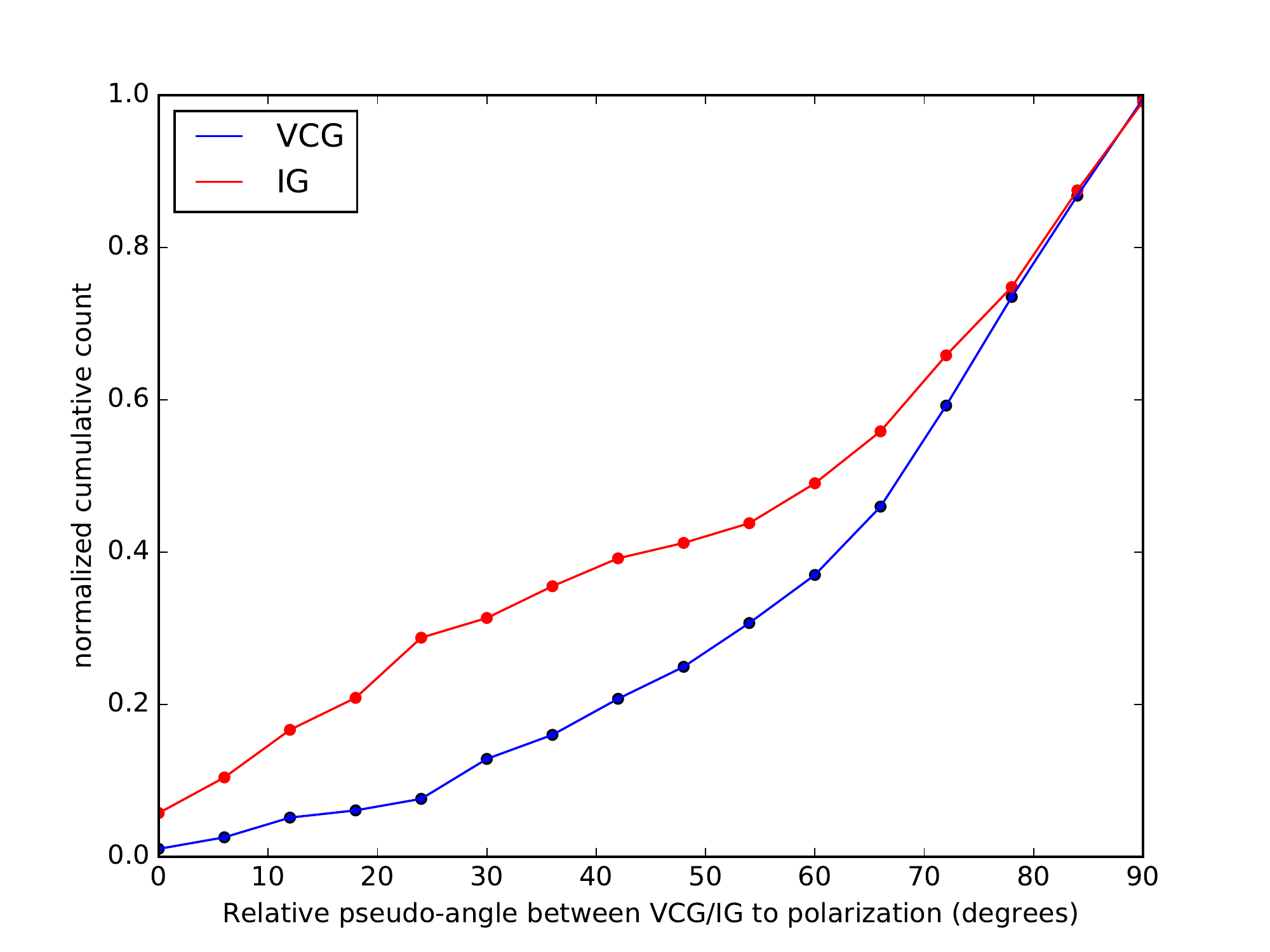}
\caption{\label{fig:gadfi-hi-hist} A cumulative histogram showing the relative pseudo-angle between sub-block averaged VCG and intensity gradient (IG) to polarization. To trace the magnetic field direction, we rotated gradient vectors by 90 degrees.}
\end{figure}

Following GL16 we provide a comparison with the alignment magnetic field as traced by polarization and the intensity gradients. The emission intensity of atomic is proportional to its column density. The column density gradients were shown to act as tracers of magnetic fields is \citep{Soler2013}.  Figure \ref{fig:gadfi-hi-hist} shows the histograms of relative orientations between velocity and intensity gradient vectors to polarization. In agreement with the theoretical expectations as well as the results in GL16, our improved procedure of calculating the VCGs shows that the latter are much better aligned with polarization compared to the intensity gradients. Indeed, nearly 80\% of the VCGs are within $45^{o}$  deviation from the polarization direction compared to 61\% of the intensity gradients. 

\begin{figure*}[pt]
\centering
\includegraphics[width=0.99\textwidth]{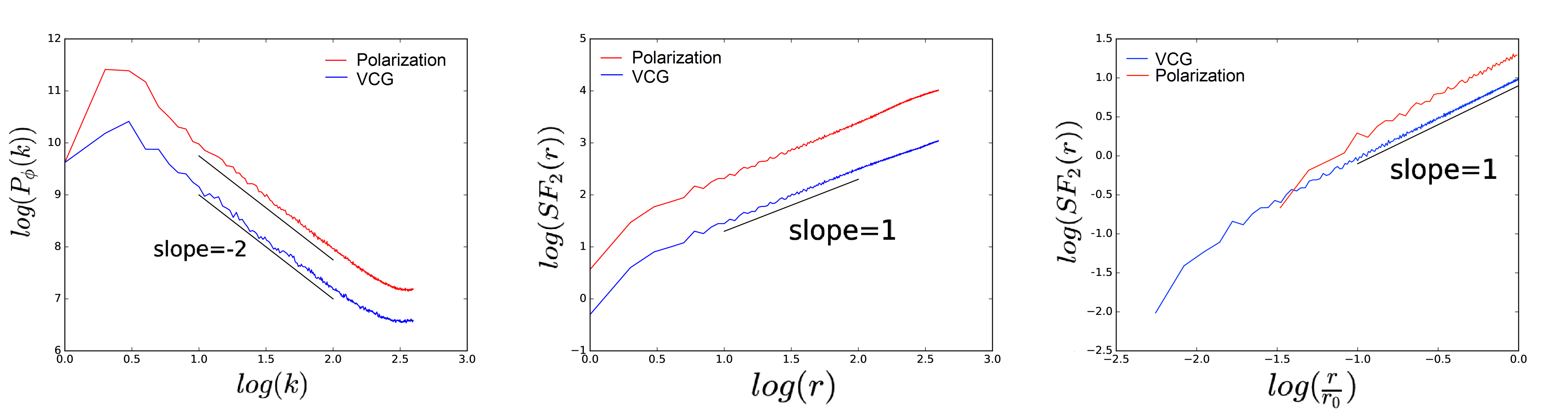}
\caption{\label{fig:vg-pol-sf} The power spectrum (Left) and second order structure function (Middle) of the sub-block averaged velocity gradient (blue) and polarization (red) from the synthetic map.. Both power spectra and structure functions show very similar behavior. Structure functions (Right) of the sub-block averaged velocity gradients and polarization angle from observation data.}
\end{figure*}
\section{Discussion}
\label{sec:discussions}
\subsection{Structure functions of velocity gradients}
The structure functions of polarization and gradient fields can also allow us to study how well-aligned they are. As the statistics of polarization are dependent on the Alfvenic Mach number $M_A$ \citep{Falceta-Goncalves2008StudiesTechnique}, the close relationship between rotated the VCGs and magnetic fields suggests that gradient statistics should have similar behavior to the polarization statistics. To compare the VCGs to polarization in synthetic maps, we extended the sub-block averaging algorithm to every point of our map, and computed the structure function in terms of the orientation $\theta$ of gradient/polarization vectors:
\begin{equation}
SF_2({\bf r}) = \langle(\theta({\bf r}')-\theta({\bf r}'+{\bf r}))^2\rangle
\end{equation}

The statistics of dust polarization are important for studying magnetic field turbulence \citep{Falceta-Goncalves2008StudiesTechnique} and for cleaning the CMB polarization maps. If we want to do the same using VCGs,
it is important test to what extent the statistics of the VCGs are similar to those revealed by polarization.
The left and the middle panels of  \ref{fig:vg-pol-sf} show the power spectra $P_\phi(k)$ and second order structure functions $SF_2({\bf r})$, respectively, of the VCGs orientation and the polarization angle. In terms of the spectra, both VCGs orientations and polarization angles exhbitit a $-2$ slope.  We also examined the structure functions for polarization and the VCG distributions from the observation data using the same procedure.  The right panel of Figure \ref{fig:vg-pol-sf} shows the structure function computed using observation data, the $+1$ slope also emerged.

\subsection{Comparison with other techniques and earlier papers}

This paper presents the first application of the VCGs to observational data arising from diffuse media. By comparing the results obtained with the VCGs and PLANCK polarimetry data, we have demonstrated the practical utility of the VCG for tracing of magnetic fields and obtaining statistical information about magnetic field in this diffuse region. 

The gradient techniques have big advantage over other techniques for estimating magnetic field direction and strengths: These techniques only require an easily available centroid. Unlike the {\it PLANCK} map, the VCG maps do not require unique multi-billion dollar satellites but can be routinely obtained with the existing spectroscopic surveys. By using different species, one can distinguish and study separately different regions along the line of sight. Combining the VCGs that trace magnetic fields in diffuse gas with polarimetry, e.g. ALMA polarimetry, that traces magnetic fields in molecular clouds, one can study what is happening with magnetic fields as star formation takes place. This may be a way to test different predictions, e.g. the prediction of magnetic flux removal through the reconnection diffusion process \citep{Lazarian2005AstrophysicalFormation,Lazarian2014ReconnectionFormation,Lazarian2012MagnetizationDiffusion}.

The alignment of density gradients were previously explored by \cite{Soler2013}. The alignment of these gradients with magnetic field is also due to the properties of turbulence. For instance, \cite{Beresnyak2005DensityTurbulencec} showed that GS95 turbulence can in some situations imprint its structure on density. However, density does not trace turbulence as directly as velocity does. Therefore, we expect more deviations of density gradients from the magnetic field direction compared to the velocity gradients. Our study confirms the conclusions in GL16 that the VCGs provide a better tracer. We expect that the density gradients are related to the filaments which align with magnetic fields as reported in \cite{Clark2015NeutralForegrounds}. Therefore we expect that the VCGs trace magnetic fields better than the filaments. 

We, however, have to stress that this region is only a particular example on how VCG works, which does not represent it is applicable everywhere without cautions on the limitations. One should understand that both density and velocity properties are important components of MHD turbulent cascades. Therefore, the deviations of the gradients from the magnetic field direction are informative. For instance, we observe an a different behavior of VCGs and density gradients in the regions of strong shocks as well as in self-gravitating regions (Yuen \& Lazarian, in prep.). Therefore there is important synergy of the simultaneous use of VCGs, density/intensity gradients and polarimetry. Adding to the list the newly suggested technique of synchrotron intensity gradients that is discussed in a new paper by Lazarian et al. (2017) increases the wealth of the available tools. This opens new ways of exploring magnetic fields in the multi-phase ISM.

We would also like to point out that while the polarimetry directions in Figure \ref{fig:gadfi-hi-vgbb} seem to be well aligned over significant patches of the sky, this does not mean that there is no turbulence there. The correspondence of the VCGs and polarization directions can be understood only if the media is turbulent. The power law behavior of the statistics related to both the VCGs and polarization directions confirms this. The fact that the power law does not correspond to the GS95 slope is due to the effects of the emitting region geometry as it discussed in \cite{Cho2002CompressiblePlasmasb,Cho2009SimulationsTurbulence}.

\section{Conclusions}
\label{sec:conclusions}

Our work provide a promising example on how the Velocity Centroid Gradient (VCG) technique introduced in GL16 traces magnetic fields in interstellar media. In the paper: 
\begin{enumerate}[wide, labelwidth=!, labelindent=0pt]
\item We provide a new robust prescription for calculating the VCGs and test this new approach using the synthetic data obtained with MHD simulations.
\item We show that with the new prescription the estimates of magnetic field strength based on the C-F approach can be improved. 
\item We apply the VCGs to the available high latitude HI GALFA data and demonstrate an excellent alignment of the direction of the VCGs and those measured by {\it PLANCK} polarization.
\item We show that the statistics of the fluctuations measured by the VCGs and polarization have the same slope for both synthetic and observational data, which suggests that VCGs could potentially be promising tool for accounting for polarized foregrounds within CMB studies.
\item The differences between the directions defined by the polarization, the VCGs and the intensity gradients carry information about the turbulent interstellar medium and this calls for the synergetic use of the three measures.

\end{enumerate}

We thank Susan Clark for her help with GALFA data. We thank Avi Loeb and Diego F. Gonzalez-Casanova useful discussions. We also thank Paul Law for his generous help with {\it PLANCK} data. The stay of KHY at UW-Madison is supported by the Fulbright-Lee Hysan research fellowship and Department of Physics, CUHK. AL acknowledges the support the NSF grant AST 1212096, NASA grant NNX14AJ53G as well as a distinguished visitor PVE/CAPES appointment at the Physics Graduate Program of the Federal University of Rio Grande do Norte, the INCT INEspao and Physics Graduate Program/UFRN.

%\bibliographystyle{yahapj}
%\bibliography{Mendeley}

\begin{thebibliography}{}
\providecommand\natexlab[1]{#1}
\providecommand\JournalTitle[1]{#1}
\bibitem[{Adam {et~al.}(2016)Adam, Ade, Aghanim, Arnaud, Ashdown, Aumont,
  Baccigalupi, Banday, Barreiro, Bartolo, Battaner, Benabed, Beno{\^{i}}t,
  Benoit-L{\'{e}}vy, Bernard, Bersanelli, Bertincourt, Bielewicz, Bock,
  Bonavera, Bond, Borrill, Bouchet, Boulanger, Bucher, Burigana, Calabrese,
  Cardoso, Catalano, Challinor, Chamballu, Chiang, Christensen, Clements,
  Colombi, Colombo, Combet, Couchot, Coulais, Crill, Curto, Cuttaia, Danese,
  Davies, Davis, de~Bernardis, de~Rosa, de~Zotti, Delabrouille, Delouis,
  D{\'{e}}sert, Diego, Dole, Donzelli, Dor{\'{e}}, Douspis, Ducout, Dupac,
  Efstathiou, Elsner, En{\ss}lin, Eriksen, Falgarone, Fergusson, Finelli,
  Forni, Frailis, Fraisse, Franceschi, Frejsel, Galeotta, Galli, Ganga, Ghosh,
  Giard, Giraud-H{\'{e}}raud, Gjerl{\o}w, Gonz{\'{a}}lez-Nuevo, G{\'{o}}rski,
  Gratton, Gruppuso, Gudmundsson, Hansen, Hanson, Harrison,
  Henrot-Versill{\'{e}}, Herranz, Hildebrandt, Hivon, Hobson, Holmes,
  Hornstrup, Hovest, Huffenberger, Hurier, Jaffe, Jaffe, Jones, Juvela,
  Keih{\"{a}}nen, Keskitalo, Kisner, Kneissl, Knoche, Kunz, Kurki-Suonio,
  Lagache, Lamarre, Lasenby, Lattanzi, Lawrence, Le~Jeune, Leahy, Lellouch,
  Leonardi, Lesgourgues, Levrier, Liguori, Lilje, Linden-V{\o}rnle,
  L{\'{o}}pez-Caniego, Lubin, Mac{\'{i}}as-P{\'{e}}rez, Maggio, Maino,
  Mandolesi, Mangilli, Maris, Martin, Mart{\'{i}}nez-Gonz{\'{a}}lez, Masi,
  Matarrese, McGehee, Melchiorri, Mendes, Mennella, Migliaccio, Mitra,
  Miville-Desch{\^{e}}nes, Moneti, Montier, Moreno, Morgante, Mortlock, Moss,
  Mottet, Munshi, Murphy, Naselsky, Nati, Natoli, Netterfield,
  N{\o}rgaard-Nielsen, Noviello, Novikov, Novikov, Oxborrow, Paci, Pagano,
  Pajot, Paoletti, Pasian, Patanchon, Pearson, Perdereau, Perotto, Perrotta,
  Pettorino, Piacentini, Piat, Pierpaoli, Pietrobon, Plaszczynski,
  Pointecouteau, Polenta, Pratt, Pr{\'{e}}zeau, Prunet, Puget, Rachen,
  Reinecke, Remazeilles, Renault, Renzi, Ristorcelli, Rocha, Rosset, Rossetti,
  Roudier, Rusholme, Sandri, Santos, Sauv{\'{e}}, Savelainen, Savini, Scott,
  Seiffert, Shellard, Spencer, Stolyarov, Stompor, Sudiwala, Sutton, Suur-Uski,
  Sygnet, Tauber, Terenzi, Toffolatti, Tomasi, Tristram, Tucci, Tuovinen,
  Valenziano, Valiviita, Van~Tent, Vibert, Vielva, Villa, Wade, Wandelt,
  Watson, Wehus, Yvon, Zacchei, \& Zonca}]{Adam2016iPlanck/iResults}
Adam, R., Ade, P. A.~R., Aghanim, N., {et~al.} 2016,
  \href{http://dx.doi.org/10.1051/0004-6361/201525820}{\JournalTitle{Astronomy {\&} Astrophysics}, 594, A8}

\bibitem[{Andersson {et~al.}(2015)Andersson, Lazarian, \&
  Vaillancourt}]{Andersson2015InterstellarAlignment}
Andersson, B.-G., Lazarian, A., \& Vaillancourt, J.~E. 2015,
  \href{http://dx.doi.org/10.1146/annurev-astro-082214-122414}{\JournalTitle{Annu.
  Rev. Astron. Astrophys}, 53, 501}

\bibitem[{Armstrong {et~al.}(1995)Armstrong, Rickett, \&
  Spangler}]{Armstrong1995ElectronMedium}
Armstrong, J.~W., Rickett, B.~J., \& Spangler, S.~R. 1995,
  \href{http://dx.doi.org/10.1086/175515}{\JournalTitle{The Astrophysical
  Journal}, 443, 209}
   
\bibitem[Beresnyak(2014)]{2014ApJ...784L..20B} Beresnyak, A.\ 2014, \apjl, 784, L20
\bibitem[{Beresnyak \& {Andrey}(2014)}]{Beresnyak2014SpectraSimulations}
Beresnyak, A., \& {Andrey}. 2014,
  \href{http://dx.doi.org/10.1088/2041-8205/784/2/L20}{\JournalTitle{The
  Astrophysical Journal Letters, Volume 784, Issue 2, article id. L20, 5 pp.
  (2014).}, 784}

\bibitem[{Beresnyak \& Lazarian(2010)}]{Beresnyak2010ScalingTurbulence}
Beresnyak, A., \& Lazarian, A. 2010,
  \href{http://dx.doi.org/10.1088/2041-8205/722/1/L110}{\JournalTitle{The
  Astrophysical Journal Letters, Volume 722, Issue 1, pp. L110-L113 (2010).},
  722, L110}

\bibitem[{Beresnyak {et~al.}(2005)Beresnyak, Lazarian, \&
  Cho}]{Beresnyak2005DensityTurbulencec}
Beresnyak, A., Lazarian, A., \& Cho, J. 2005,
  \href{http://dx.doi.org/10.1086/430702}{\JournalTitle{The Astrophysical
  Journal, Volume 624, Issue 2, pp. L93-L96.}, 624, L93}

\bibitem[{Boldyrev(2006)}]{Boldyrev2006SpectrumTurbulence}
Boldyrev, S. 2006,
  \href{http://dx.doi.org/10.1103/PhysRevLett.96.115002}{\JournalTitle{Physical
  Review Letters}, 96, 115002}

\bibitem[{Brandenburg \&
  Lazarian(2013)}]{Brandenburg2013AstrophysicalTurbulence}
Brandenburg, A., \& Lazarian, A. 2013,
  \href{http://dx.doi.org/10.1007/s11214-013-0009-3}{\JournalTitle{Space
  Science Reviews, Volume 178, Issue 2-4, pp. 163-200}, 178, 163}

\bibitem[{Chandrasekhar \& Fermi(1953)}]{Chandrasekhar1953ProblemsField.}
Chandrasekhar, S., \& Fermi, E. 1953,
  \href{http://dx.doi.org/10.1086/145732}{\JournalTitle{The Astrophysical
  Journal}, 118, 116}

\bibitem[{Chepurnov \& Lazarian(2010)}]{Chepurnov2010ExtendingData}
Chepurnov, A., \& Lazarian, A. 2010,
  \href{http://dx.doi.org/10.1088/0004-637X/710/1/853}{\JournalTitle{The
  Astrophysical Journal, Volume 710, Issue 1, pp. 853-858 (2010).}, 710, 853}

\bibitem[{Cho \& Lazarian(2002)}]{Cho2002CompressiblePlasmasb}
Cho, J., \& Lazarian, A. 2002,
  \href{http://dx.doi.org/10.1103/PhysRevLett.88.245001}{\JournalTitle{Physical
  Review Letters, vol. 88, Issue 24, id. 245001}, 88}

\bibitem[{Cho \& Lazarian(2003)}]{Cho2003CompressibleImplicationsb}
---. 2003,
  \href{http://dx.doi.org/10.1046/j.1365-8711.2003.06941.x}{\JournalTitle{Monthly
  Notices of the Royal Astronomical Society, Volume 345, Issue 12, pp.
  325-339.}, 345, 325}

\bibitem[{Cho \& Lazarian(2009)}]{Cho2009SimulationsTurbulence}
---. 2009,
  \href{http://dx.doi.org/10.1088/0004-637X/701/1/236}{\JournalTitle{The
  Astrophysical Journal, Volume 701, Issue 1, pp. 236-252 (2009).}, 701, 236}

\bibitem[{Cho {et~al.}(2001)Cho, Lazarian, \&
  Vishniac}]{Cho2001SimulationsMedium}
Cho, J., Lazarian, A., \& Vishniac, E. 2001,
  \href{http://dx.doi.org/10.1086/324186}{\JournalTitle{The Astrophysical
  Journal, Volume 564, Issue 1, pp. 291-301.}, 564, 291}

\bibitem[{Cho \& Vishniac(2000)}]{Cho2000TheTurbulence}
Cho, J., \& Vishniac, E.~T. 2000,
  \href{http://dx.doi.org/10.1086/309213}{\JournalTitle{The Astrophysical
  Journal, Volume 539, Issue 1, pp. 273-282.}, 539, 273}

\bibitem[{Clark {et~al.}(2015)Clark, Hill, Peek, Putman, \&
  Babler}]{Clark2015NeutralForegrounds}
Clark, S.~E., Hill, J.~C., Peek, J. E.~G., Putman, M.~E., \& Babler, B.~L.
  2015,
  \href{http://dx.doi.org/10.1103/PhysRevLett.115.241302}{\JournalTitle{Physical
  Review Letters}, 115, 1}

\bibitem[{Draine(2011)}]{Draine2011PhysicsMedium}
Draine, B.~T. 2011, {Physics of the interstellar and intergalactic medium}
  (Princeton University Press), 540

\bibitem[{Falceta-Goncalves {et~al.}(2008)Falceta-Goncalves, Lazarian, \&
  Kowal}]{Falceta-Goncalves2008StudiesTechnique}
Falceta-Goncalves, D., Lazarian, A., \& Kowal, G. 2008,
  \href{http://dx.doi.org/10.1086/587479}{\JournalTitle{The Astrophysical
  Journal, Volume 679, Issue 1, article id. 537-551, pp. (2008).}, 679}

\bibitem[{Goldreich(1995)}]{GoldreichP.Sridhar1995GS95IITurbulence}
Goldreich, P.~ \& Sridhar, S. 1995,
  \href{http://dx.doi.org/10.1086/174600}{\JournalTitle{The Astronomical
  Journal}, 438, 763}

\bibitem[{{Gonz{\'a}lez-Casanova} \& {Lazarian}(2016)}]{2016arXiv160806867G}
{Gonz{\'a}lez-Casanova}, D.~F., \& {Lazarian}, A. 2016, \JournalTitle{ArXiv
  e-prints}, \href{http://arxiv.org/abs/1608.06867}{{\sffamily
  arXiv:1608.06867}}

\bibitem[{Hayes {et~al.}(2006)Hayes, Norman, Fiedler, Bordner, Li, Clark, \&
  Low}]{Hayes2006}
Hayes, J.~C., Norman, M.~L., Fiedler, R.~A., {et~al.} 2006, 188

\bibitem[{Kowal \& Lazarian(2010)}]{Kowal2010VelocityScalingsb}
Kowal, G., \& Lazarian, A. 2010,
  \href{http://dx.doi.org/10.1088/0004-637X/720/1/742}{\JournalTitle{The
  Astrophysical Journal, Volume 720, Issue 1, pp. 742-756 (2010).}, 720, 742}

\bibitem[{Lazarian \& {A.}(2005)}]{Lazarian2005AstrophysicalFormation}
Lazarian, A. 2005,
  \href{http://dx.doi.org/10.1063/1.2077170}{\JournalTitle{MAGNETIC FIELDS IN
  THE UNIVERSE: From Laboratory and Stars to Primordial Structures. AIP
  Conference Proceedings, Volume 784, pp. 42-53 (2005).}, 784, 42}
\bibitem[{{Lazarian}(2007)}]{2007JQSRT.106..225L}
{Lazarian}, A. 2007,
  \href{http://dx.doi.org/10.1016/j.jqsrt.2007.01.038}{\JournalTitle{\jqsrt},
  106, 225}


\bibitem[{Lazarian \& {A.}(2014)}]{Lazarian2014ReconnectionFormation}
---. 2014,
  \href{http://dx.doi.org/10.1007/s11214-013-0031-5}{\JournalTitle{Space
  Science Reviews}, 181, 1}

\bibitem[{Lazarian {et~al.}(2012)Lazarian, Esquivel, \&
  Crutcher}]{Lazarian2012MagnetizationDiffusion}
Lazarian, A., Esquivel, A., \& Crutcher, R. 2012,
  \href{http://dx.doi.org/10.1088/0004-637X/757/2/154}{\JournalTitle{The
  Astrophysical Journal, Volume 757, Issue 2, article id. 154, 20 pp. (2012).},
  757}

\bibitem[{Lazarian \& Vishniac(1999)}]{Lazarian1999ReconnectionField}
Lazarian, A., \& Vishniac, E.~T. 1999,
  \href{http://dx.doi.org/10.1086/307233}{\JournalTitle{The Astrophysical
  Journal, Volume 517, Issue 2, pp. 700-718.}, 517, 700}
  
%\bibitem[Lazarian, Yuen et al. (2017}]{Lazarian2017synchrotron} Lazarian, A., Yuen, K. H., Lee, H. \& Cho, J.
%2017, ApJ, submitted

\bibitem[{Li {et~al.}(2014)Li, Banerjee, Pudritz, J{\o}rgensen, Shang,
  Krasnopolsky, \& Maury}]{Li2014a}
Li, Z.-Y., Banerjee, R., Pudritz, R.~E., {et~al.} 2014,
  \href{http://adsabs.harvard.edu/abs/2014arXiv1401.2219L}{\JournalTitle{arXiv.org},
  1401, 2219}

\bibitem[{Lithwick \& Goldreich(2001)}]{Lithwick2001CompressiblePlasmas}
Lithwick, Y., \& Goldreich, P. 2001,
  \href{http://dx.doi.org/10.1086/323470}{\JournalTitle{The Astrophysical
  Journal, Volume 562, Issue 1, pp. 279-296.}, 562, 279}

\bibitem[{Maron \& Goldreich(2000)}]{Maron2000SimulationsTurbulence}
Maron, J., \& Goldreich, P. 2000,
  \href{http://dx.doi.org/10.1086/321413}{\JournalTitle{The Astrophysical
  Journal, Volume 554, Issue 2, pp. 1175-1196.}, 554, 1175}

\bibitem[{Norman(2000)}]{Norman2000}
Norman, M.~L. \href{http://arxiv.org/abs/astro-ph/0005109}{2000, 6, 6}

\bibitem[{Ostriker {et~al.}(2000)Ostriker, Stone, \& Gammie}]{Ostriker2000}
Ostriker, E.~C., Stone, J.~M., \& Gammie, C.~F.
  \href{http://dx.doi.org/10.1086/318290}{2000, 56}
\bibitem[Otto et al.(2017)]{Otto17} Otto, F., Ji, W., \& Li, H.-b.\ 2017, arXiv:1701.01806 
\bibitem[{Peek {et~al.}(2011)Peek, Heiles, Douglas, Lee, Grcevich,
  Stanimirovic, Putman, Korpela, Gibson, Begum, Saul, Robishaw, \&
  Krco}]{Peek2011The1}
Peek, J. E.~G., Heiles, C., Douglas, K.~A., {et~al.} 2011,
  \href{http://dx.doi.org/10.1088/0067-0049/194/2/20}{\JournalTitle{The
  Astrophysical Journal Supplement, Volume 194, Issue 2, article id. 20, 13 pp.
  (2011).}, 194}

\bibitem[{Pillai {et~al.}(2015)Pillai, Kauffmann, Tan, Goldsmith, Carey, \&
  Menten}]{Pillai2015}
Pillai, T., Kauffmann, J., Tan, J.~C., {et~al.}
  \href{http://dx.doi.org/10.1088/0004-637X/799/1/74}{2015, 74}

\bibitem[{Soler {et~al.}(2013)Soler, Hennebelle, Martin,
  Miville-Desch{\^{e}}nes, Netterfield, \& Fissel}]{Soler2013}
Soler, J.~D., Hennebelle, P., Martin, P.~G., {et~al.}
  \href{http://dx.doi.org/10.1088/0004-637X/774/2/128}{2013, 16}

\bibitem[{Zhang {et~al.}(2014)Zhang, Qiu, Girart, {Hauyu}, {Liu}, Tang, Koch,
  Li, Keto, Ho, Rao, Lai, Ching, Frau, Chen, Li, Padovani, Bontemps, Csengeri,
  \& Juarez}]{Zhang2014}
Zhang, Q., Qiu, K., Girart, J.~M., {et~al.}
   \href{http://arxiv.org/abs/1407.3984}{2014, 1}


\end{thebibliography}

\end{document}